\documentstyle{article}
\begin{document}
\rightline{Nuovo Cimento B 112, 1191-1192 (Aug. 1997)}
\begin{center}
{\bf Technological picture of the Planck scale universe}\\

H. Rosu
\footnote{e-mail: rosu@ifug.ugto.mx\\
%Additional quite old address: IGSS, Magurele-Bucharest, Romania
}\\[2mm]

Instituto de F\'{\i}sica - IFUG,
Apdo Postal E-143, 37150 Le\'on, Gto, M\'exico\\
\end{center}

%\date{December 2/96; June 25/97}

%{\baselineskip=20pt
%\begin{center}
%{\bf Abstract}

{\bf Summary.} - A picture of the Planck scale universe as a network of
mesoscopic (with respect to that scale) resonators connected through
quasi-one-dimensional waveguides is presented.

\bigskip

%\end{center}

PACS 04.60 - Quantum theory of gravitation

\bigskip
\bigskip

As is well known there is no theory of boundary conditions in quantum
cosmology. In the literature one can find only reasonable proposals.
Hartle-Hawking no-boundary and Vilenkin tunneling conditions are the most
used and known ones.
I have been confronted with the problem of boundary conditions in
recent works on the supersymmetric
double Darboux method for a couple of simple minisuperspace models \cite{r}.
In the one-dimensional supersymmetric nonrelativistic context, this
procedure means, mathematically speaking, to work with the general Riccati
solution (general Witten superpotential), which  generates a one-parameter
family of strictly isospectral cosmological potentials. This strictly
isospectral family of potentials depends on the wavefunction, and as such,
quite strong
deformations are introduced. Roughly speaking, in the single
Darboux procedure as applied to quantum cosmology, one goes from zero
energy to a ``fermionic" (scattering) level. In the double
Darboux method one returns to zero energy to find a modified
nonlinear Wheeler-DeWitt (WDW) equation.

The interesting point is that
one can work with the general superposition solution of the WDW equation
to obtain isospectral modulational
effects directly on it. The strictly isospectral states have coherent
(localization) properties. For example, in the simpler case of
nonrelativistic quantum
mechanics the double Darboux technique may lead to bound states in the
continuum \cite{psp}. Thus, one can claim that the cosmological isospectral
``wavefunctions" are nonequilibrium states that can be maintained by the
coupling to a general quantum reservoir (possibly viewed as a collection
of sets of reservoirs) produced by the neglected degrees of
freedom in minisuperspace models and making up the cosmological
``environment".

In the recent work on Whittaker quantum universes \cite{w} I concluded
that the multitude of strictly isospectral WDW superposition solutions
are in fact modes of a set of
both stable and unstable resonators. But how to think of resonators at
the Planck scale ? The RMP colloquium on string duality by
Polchinski \cite{pol} suggested me one way of doing this.
When Polchinski discusses
higher dimensions he introduces a simple waveguide model for the
confinement of more dimensions. Actually, the Planck scale waveguides of
cross section $l_{P}^{2}$ in $\hbar=c=1$ units may be thought as various
types of strings.
On the other hand, the mesoscopic resonators are chunks of more or less
well-defined four dimensional spacetimes.
Matter and radiation are circulating within the resonator regions and along
the Planck waveguides.

This picture of the Planck scale universe might be thought of as a consequence
of the double Darboux method as I will argue in the following. As I said
the method introduces some coherence effects in the scattering states.
To better see the connection I am thinking of, I shall now describe what
happens in nanoelectronics where
precisely such a type of states are very important.
They are called current-carrying
states, belong to an effective-mass Schr\"odinger equation, and are related
to the main discovery in the technology at that scale, which is coherent
ballistic transport in narrow channels formed by imposing a potential on a
two-dimensional electron gas. This coherent nanoelectronic transport
suggests a device structure made of a resonator connected through small
apertures to quantum waveguides \cite{pp}.
Moreover, Lent and Kirkner \cite{lk} introduced so-called quantum
transmitting boundary
conditions for the current-carrying states that can be implemented in
quantum cosmology as well. Essentially, the boundary conditions are
formulated in terms of a functional relating the value of the wave
function's normal derivative at a particular point to the values of the wave
function at all the other points along the waveguide boundaries
(resonator apertures).

In conclusion, I have briefly discussed a mesoscopic device analogy for the
Planck scale universe. It may help to introduce new types of boundary
conditions in quantum cosmology and to find reasons for the
cosmological coherence at large scales.

\begin{center}  ***  \end{center}

Work supported in part by CONACyT Project No. 4868-E9406.

%%%%%%%%%%%%%%%%%%%%%%%%%%%%%%%%%%%%%%%%%%%

\end{document}